\documentclass[ 11pt]{article}

\usepackage{graphicx}
\usepackage{indentfirst}
\usepackage{amsmath, amsthm, amssymb}
\usepackage[usenames,dvipsnames]{color}
\usepackage{fullpage}
\usepackage{natbib}
\usepackage{setspace}
\setcitestyle{comma}
\usepackage{epstopdf}
\usepackage[labelfont=bf,labelsep=period,justification=raggedright]{caption}

\newcommand{\e}{{\rm e}}
\renewcommand{\d}{{\rm d}}
\newcommand{\pd}{\partial}

\allowdisplaybreaks[1]

\newcommand{\ignore}[1]{}

\begin{document}

% Title must be 150 characters or less
\begin{flushleft}

{\Large
\textbf{Optimizing working memory with heterogeneity of recurrent cortical excitation}
}

\vspace{0.75 cm}
% Insert Author names, affiliations 

Zachary P. Kilpatrick$^{1}$, 
Bard Ermentrout$^{2,3}$, 
Brent Doiron$^{2,3}$
\\
\vspace{0.25 cm}

{\footnotesize
{1}: Department of Mathematics, University of Houston, Houston, TX, USA.\\
{2}: Department of Mathematics, University of Pittsburgh, Pittsburgh, PA, USA.\\
{3}: Center for the Neural Basis of Cognition, Pittsburgh, PA, USA.\\ 
}

%J. Neurosci. information
\vspace{.75cm}
Journal section: Behavioral/Systems/Cognitive \\
Corresponding author: \\ 
\hspace{.5cm} Zachary P. Kilpatrick \\
\hspace{.5cm} Department of Mathematics, University of Houston\\
\hspace{.5cm} Phillip G. Hoffman 651\\
\hspace{.5cm} Houston, TX 77204-3008 \\
\hspace{.5cm} email: zpkilpat@math.uh.edu \\ 
Number of figures: 9 \\
Number of tables: 0 \\
Pages: 42 \\
Word counts: Abstract - 204, Introduction - 500, Discussion - 1481  \\
Conflict of interest: We declare no competing interests. \\ 
\section*{Acknowledgments}
We thank Robert Rosenbaum, Ashok Litwin-Kumar, and Kre\v{s}imir Josi\'{c} for useful discussions.  Funding was provided by NSF-DMS-1121784 (B.D.), NSF-DMS-1311755 (Z.P.K.); and NSF-DMS-1219753 (B.E.).
\doublespacing

\end{flushleft}

\newpage

\clearpage
\doublespacing 
\section*{Abstract}

A neural correlate of parametric working memory is a stimulus specific rise in neuron firing rate that persists long after the stimulus is removed. Network models with local excitation and broad inhibition support persistent neural activity, linking network architecture and parametric working memory.  Cortical neurons receive noisy input fluctuations which causes persistent activity to diffusively wander about the network, degrading memory over time.  We explore how cortical architecture that supports parametric working memory affects the diffusion of persistent neural activity.  Studying both a spiking network and a simplified potential well model, we show that spatially heterogeneous excitatory coupling stabilizes a discrete number of persistent states, reducing the diffusion of persistent activity over the network.  However, heterogeneous coupling also coarse-grains the stimulus representation space, limiting the capacity of parametric working memory.  The storage errors due to coarse-graining and diffusion tradeoff so that information transfer between the initial and recalled stimulus is optimized at a fixed network heterogeneity. For sufficiently long delay times, the optimal number of attractors is less than the number of possible stimuli, suggesting that memory networks can under-represent stimulus space to optimize performance.  Our results clearly demonstrate the effects of network architecture and stochastic fluctuations on parametric memory storage.                

\pagestyle{plain}

\section{Introduction}
\label{intro}

Persistent neural activity occurs in prefrontal \citep{furster73,funahashi89,romo99} and parietal \citep{pesaran02} cortex during the retention interval of parametric working memory tasks. Model networks of stimulus tuned neurons that are connected with local slow excitation  \citep{wang99} and broadly tuned inhibitory feedback \citep{compte00,goldmanrakic95} exhibit localized and persistent high rate spike train patterns called ``bump" states  \citep{compte00,renart03}. Bumps have initial locations that are stimulus-dependent,
%and since each stimulus recruits an activity bump with a unique initial location
so population activity provides a code for the remembered stimulus \citep{durstewitz00}. These models relate cortical architecture to persistent neural activity, and are a popular framework for studying working memory \citep{wang01,brody03}.  

Neural variability is present in all brain regions and limits neural coding in many sensory, motor, and cognitive tasks \citep{stein05,faisal08,laing09}.  In parametric working memory networks, dynamic input fluctuations cause bump states to wander diffusively \citep{compte00,laing01,wu08,polk12,burak12,kilpatrick13}, degrading stimulus storage over time.  Psychophysical data shows that the spread of the recalled position increases with delay time \citep{white94,ploner98}, consistent with diffusive wandering of a bump state.  While several results examine how bump formation depends upon neural architecture, little is known about how cortical wiring affects the diffusion of persistent neural activity.

The response properties of cells are often heterogeneous \citep{ringach02}, a feature that can improve population-based codes \citep{chelaru08,shamir08,marsat10,osborne08,padmanabhan10}.  In particular, there is a large degree of variation in synaptic plasticity and cortical wiring in prefrontal cortical networks involved in persistent activity during working memory tasks \citep{rao99,wang06}.  Heterogeneity in excitatory coupling quantizes the neural space used to store inputs, reducing the network's overall storage capacity \citep{renart03,itskov11}. On the other hand, stabilizing a discrete number of network states improves the robustness of working memory dynamics to parameter perturbation \citep{rosen72,koulakov02,goldman03,brody03,miller06}.  In this study we investigate how stabilization introduced by synaptic heterogeneity affects the temporal diffusion of persistent neural activity.  

%First, the fragile nature of line attractor networks requires careful tuning of recurrent synaptic connections for proper memory storage \citep{renart03,itskov11}.  Integrator networks, which accumulate information evidence or input serially in time, also typically exploit a line attractor for their computations and demonstrate similar fragility \citep{goldman03}. One solution to this problem is to consider networks of bistable neurons \citep{aksay03}. This enhances the robustness of integrators by replacing the underlying line attractor with a chain of discrete stable states \citep{rosen72,koulakov02,goldman03,brody03,miller06}. These network models display integration that is very robust to parameter variations, circumnavigating the criticism.  The cost of this robustness is the loss of the accuracy that continuum attractors provide in coding parametric stimuli. We propose that spatial heterogeneity in synaptic projections \citep{rao99,lewis00} or single neuron properties \citep{zaitsev05} can produce similar robustness in bump attractor networks.   %However, since delay-response tasks often present stimuli at a finite number (e.g. eight) of spatial locations \citep{funahashi89}. Thus, a discrete network structure could arise due to plasticity mechanisms during training \citep{wang08} and not compromise the stimulus coding of the already discrete stimulus space.   

We show that spatial heterogeneities in the excitatory architecture of a spiking network model of working memory reduce the rate with which bumps diffuse away from their initial position.  However, the same heterogeneities limit the number of stable network states used to store memories.   
%Using a simplified potential well model, we directly calculate an effective diffusion coefficient for the bump in terms of the spatial heterogeneity of coupling.   Information theoretic methods uncover a tradeoff between the quantization error in encoding due to state discretization and the error due to diffusion during the storage period.  This tradeoff leads to an optimal number ($n_{opt}$) of attractors that best encodes one of a fixed number ($m$) of stimuli for a fixed retention time. For large $m$ and long retention times we find the counterintuitive result that $n_{opt}<m$.
A tradeoff between these consequences maximizes the transfer of stimulus information at a specific degree of network heterogeneity.  For a large number of stimulus locations and long retention times we show that network architectures that under-represent stimulus space can optimize performance in working memory tasks.

\section{Materials and Methods}

\noindent
{\em Recurrent network architecture.} We employed a ring architecture for our network, commonly used for generating persistent activity to represent direction between $0^{\circ}$ and $360^{\circ}$ \citep{benyishai95,compte00} with $N_E = 256$ pyramidal cells ($E$) and $N_I = 64$ interneurons ($I$). Each leaky integrate-and-fire neuron \citep{laing01} was distinguished by its cue orientation preference $\theta_j$, where $\theta_j(E) = \Delta_E \cdot j$ ($j=1,...,N_E$) and $\theta_j(I) = \Delta_I \cdot j$ ($j=1,...,N_I$) for $\Delta_E = 360/256$ and $\Delta_I = 360/64$, respectively. The subthreshold membrane potential of each neuron, $V_{\alpha} ( \theta_j, t)$ ($\alpha = E,I$), obeyed
\begin{align*}
\frac{\d V_{\alpha} ( \theta_j , t)}{\d t} & = - V_{\alpha} ( \theta_j, t) + I_{\alpha} + I_{ext, \alpha} (\theta_j, t)+ I_{syn, \alpha}( j , t ) + I_{n, \alpha} (t) , \ \ \ \  \alpha = E,I,
\end{align*}
where $I_E = 0.6$ and $I_I = 0.6$ are bias currents that determine the resting potential of $E$ and $I$ neurons. The external current
\begin{align*}
I_{ext, E}(\theta_j, t) = I_0 \exp \left[ - \left( \frac{\theta_j - \theta_{ext}}{I_d} \right)^2 \right], \ \ t \in [T_{\rm ON},T_{\rm OFF}]; \ \ \ \ \  I_{ext,E}(\theta_j,t) = 0, \ \ {\rm otherwise},
\end{align*}
represents sensory input received only by pyramidal neurons, where $I_0 = 2$ is the input amplitude, $I_d = 3$ determines input width, and $\theta_{ext}$ is the cue position. The stimulus was turned on at $T_{ON} = -1$s and off at $T_{OFF} = 0$s. Interneurons received no external input, so $I_{ext,I} = 0$. Voltage fluctuations were represented by the white noise process $I_{n, \alpha}(t)$ with variance $\sigma_{V,\alpha}^2$ ($\sigma_{V,E} = 0.5$ and $\sigma_{V,I} = 0.3$). We scaled and nondimensionalized voltage so the threshold potential $V_{th} = 1$ and the reset potential $V_{res} = 0$ for all neurons.

Synaptic currents were mediated by a sum of AMPA, NMDA, and GABA currents:
\begin{align*}
I_{syn, \alpha} (j,t) &= I_{AMPA,\alpha}(j,t) + I_{NMDA,\alpha}(j,t) - I_{GABA,\alpha}(j,t),
\end{align*}
each modeled as
\begin{align*}
I_{AMPA, \alpha} (j,t) &= A_{AMPA,\alpha} \sum_{k=1}^{N_E} W_{AMPA,\alpha} ( \theta_j, \theta_k) s_{AMPA,E} ( \theta_k , t), \\
I_{NMDA, \alpha} (j,t) &= A_{NMDA,\alpha} \sum_{k=1}^{N_E} W_{NMDA,\alpha} ( \theta_j , \theta_k) s_{NMDA,E} ( \theta_k, t), \\
I_{GABA, \alpha} &= A_{GABA, \alpha} \sum_{k=1}^{N_I} W_{GABA, \alpha} ( \theta_j, \theta_k ) s_{GABA,I} ( \theta_k, t),
\end{align*}
where $A_{AMPA, E} = 1$, $A_{NMDA,E} = 2$, $A_{GABA,E} = 0.81$, $A_{AMPA,I} = 1$, $A_{NMDA,I} = 1$, and $A_{GABA,I} = 0$ all scaled the synaptic conductance.  Orientation preference was introduced into synaptic conductance by the spatially decaying functions (Fig. \ref{fig1}{\bf A})
\begin{align}
W_{\beta, \alpha}( \theta_j, \theta_k) = \exp \left[ \frac{ \cos ( \pi [\theta_j - \theta_k]/180 ) -1}{d_{\beta, \alpha}} \right],  \label{Whom}
\end{align}
where $d_{AMPA, E} = 0.32$, $d_{NMDA,E} = 0.32$, $d_{GABA,E} = 5$, $d_{AMPA,I} = 5$, and $d_{NMDA,I} = 5$. 
%We did not include GABA projections to interneurons in this study, as the small proportion of inhibitory-inhibitory synaptic connections expected in the prefrontal cortex would not alter our results substantially \citep{somogyi98}. 
The function Eq.~(\ref{Whom}) was used for excitatory (AMPA and NMDA) synapses between pyramidal ($E$) cells in the case of spatially homogeneous connectivity. In the case of spatially heterogeneous synaptic strength (Fig. \ref{fig2}), the strength of AMPA and NMDA connections between pyramidal cells ($E$) was given by
\begin{align*}
W_{\beta, E} ( \theta_j, \theta_k) = \left[ 1 + h \cos (\pi n \theta_k / 180) \right] \exp \left[ \frac{ \cos ( \pi [\theta_j - \theta_k]/180 ) -1}{d_{\beta, E}} \right],
\end{align*}
where $h = 0.025$ represents the strength of the heterogeneity and $n$ is the frequency of the heterogeneity, which must be integer valued. 

%Synaptic connectivity like this could come about through reward-based plasticity mechanisms during training \citep{schultz98,wang08,klingberg10}, but we did not explicitly implement such a learning rule. Spatially periodic variation in synaptic conductance is motivated by the patchy structure of pyramidal-pyramidal synaptic connections observed in prefrontal cortex \citep{levitt93,lewis00}.

Synaptic gating variables of type $\beta = AMPA, NMDA,$ or $GABA$ associated with a neuron at location $\theta_j$ were instantaneously activating and exponentially decaying as described by
\begin{align*}
\frac{\d s_{\beta} ( \theta_j, t)}{\d t} &= - \frac{s_{\beta} ( \theta_j , t)}{\tau_{\beta}} + \delta (  V_{\alpha} ( \theta_j , t) - 1) + I_{s,\beta}(t),
\end{align*}
where $\alpha = E$ for $\beta = AMPA$ and $NMDA$ while $\alpha = I$ for $\beta = GABA$. Instantaneous activation is represented here using the delta function $\delta$, so $s_{\beta}(\theta_j,t)$ increments by 1 when $V_{\alpha}(\theta_j,t)$ attains the spike threshold $V_{th}=1$. Decay time constants for each synapse type are $\tau_{AMPA} = 5$ms, $\tau_{NMDA} = 100$ms, and $\tau_{GABA} = 20$ms. Fluctuations in conductance were introduced into each synapse with the term $I_{s, \beta} (t)$, which is white noise with variance $\sigma_{s,\beta}^2$ ($\sigma_{s,AMPA} = 0.1$, $\sigma_{s,NMDA} = 0.45$, and $\sigma_{s,GABA}=0.05$). We take the variance of noise to NMDA synapses to be high $\sigma_{s,NMDA} = 0.45$ because it leads to high variances in the spike times, as commonly observed in prefrontal cortical neurons during the delay period of working memory tasks \citep{compte03}. In addition, this generates an error of a few degrees in the recall of cue position for delay periods of 2-10 seconds, as observed in psychophysical experiments \citep{white94}.

Numerical simulations were done using an Euler-Maruyama method with timestep {\tt dt = 0.1}ms and normally distributed random initial conditions. Spike time rastergrams were smoothed to generate population firing rates as a function of degree and time, whose maximum at each time were used to calculate the centroid of the bump (Figs \ref{fig1}{\bf D,E} and \ref{fig2}). Variances (Figs \ref{fig1}{\bf F} and \ref{fig3}) and probability densities (Figs \ref{fig1}{\bf E} and \ref{fig2}) were computed using 1000 values for the bump centroid across 10s. Linear fits of variance in the case of spatially homogeneous synapses and spatially heterogeneous synapses with $n=8$ and $n=4$ (Fig. \ref{fig3}) were performed using linear regression. \\

\noindent
{\em Diffusion in the potential well model.} To analyze the diffusive dynamics of the bump, we studied an idealized model of bump motion.  
%of a particle diffusing on a periodic potential. We called this the potential well model, since it tracks a single variable corresponding to the remembered location of a cue. 
%This model captures the periodic potential structure that spatial heterogeneity imposed on the attractor landscape for the bump. 
In this model the bump position $\phi(t)$ obeys the stochastic differential equation  \citep{lifson62,lindner01}
\begin{align}
\d \phi (t) = - h \sin [ n \phi (t) ] \d t + \sigma \d W(t). \label{sinsde}
\end{align}
Here $\phi(t)$ was restricted to the periodic domain $\phi \in ( - \pi, \pi ]$ and $\d W$ was a standard white noise process. 
%Our analysis was performed in the radian variable $\phi$ and converted to a degree variable $\theta$ at the end. 
The first term in Eq. \eqref{sinsde} models the periodic spatial heterogeneity that is responsible for attractor dynamics.  Heterogeneity is parametrized by its strength $h$ and spatial frequency $n$. In this framework the dynamics of $\phi(t)$ is a diffusive process occurring on an energy landscape defined by the periodic potential
\begin{align}
U_n(\phi) = - \frac{h}{n} \cos (n \phi),  \label{cospot}
\end{align}
producing $n$ attractors or stable nodes (Fig. \ref{fig4}{\bf A}). These attractors occur at the minima of Eq.~(\ref{cospot}), given by $\phi = 2 j \pi /n$ where $j=1,...,n$. They are separated from one another by repellers or saddles at the maxima of Eq.~(\ref{cospot}). %Our main reason for choosing the well height to decrease with $n$ in (\ref{cospot}) was that this formulation produced diffusive behavior that resembled the spiking model. 

To analyze the model, we reformulated Eq.~(\ref{sinsde}) as an equivalent Fokker-Planck equation \citep{risken96}
\begin{align}
\frac{\pd p(\phi,t)}{\pd t} = \frac{\pd}{\pd \phi} \left[ h \sin (n \phi) p(\phi,t) \right] + \frac{\sigma^2}{2} \frac{\pd^2 p(\phi,t)}{\pd \phi^2},  \label{fpsin}
\end{align}
where $p(\phi,t)$ is the probability density of finding the bump at a given value $\phi$ at time $t$. For ease of analytic calculations we let $\phi$ evolve on an infinite domain. Since we worked in parameter regimes where the resulting spread of probability densities was relatively narrow, this did not considerably alter the results. Also, experimentally measured errors in cue recall are typically not large enough to span more than a quarter of the possible stimulus space \citep{ploner98}. For large times and sufficiently high frequency $n$, the variance of the stochastic process Eq.~(\ref{sinsde}) can be quantified using an effective diffusion coefficient \citep{lindner01}
\begin{align}
D_{\rm eff} = \frac{1}{2} \lim_{t \to \infty} \frac{\langle \Delta \phi(t)^2 \rangle}{t} = \frac{1}{2} \lim_{t \to \infty} \frac{\langle [\phi(t) - \langle \phi(t) \rangle ]^2 \rangle}{t}.   \label{infdeff}
\end{align}
The density $\rho(\phi, t)$ tends asymptotically to
\begin{align}
p_{asy}(\phi,t) = \frac{p_0(\phi)}{\sqrt{4 \pi D_{\rm eff} t}} \exp \left[ - \frac{\phi^2}{4 D_{\rm eff} t} \right]  ,  \label{fpasym}
\end{align}
where $p_0 (\phi)$ is the stationary, periodic solution of the Fokker-Planck equation, Eq.~(\ref{fpsin}), given by 
%We computed $p_0( \phi)$ by taking the limit $t \to \infty$, so $\pd p(\phi,t)/\pd t \to 0$, in (\ref{fpsin}), yielding the ordinary differential equation
%\begin{align*}
%0 & = \frac{\d}{\d \phi} \left[ h \sin (n\phi) p_0(\phi) \right] + \frac{\sigma^2}{2} \frac{\d^2 p_0(\phi)}{\d \phi^2},
%\end{align*}
%which can be integrated to give
%\begin{align*}
%\frac{\d p_0(\phi)}{\d \phi} + \frac{2 h \sin (n\phi)}{\sigma^2} p_0(\phi) &= C.
%\end{align*}
%Using an integrating factor and bounding the solution, we found
\begin{align*}
p_0(\phi) = \chi \exp \left[ \frac{2 h \cos (n\phi)}{\sigma^2 n} \right],
\end{align*}
where $\chi$ is a normalization factor. The approximation, Eq.~(\ref{fpasym}), matches realizations of the full stochastic process Eq.~(\ref{sinsde}) very well (see Fig. \ref{fig4}{\bf C}). Clearly, the frequency of the probability distribution's microscale is commensurate with that of the periodic potential Eq.~(\ref{cospot}). We were mainly concerned with computing the effective diffusion of the stochastic process defined by Eq.~(\ref{sinsde}). Remarkably, second order statistics are still well approximated by ignoring the micro-periodicity of the density in Eq.~(\ref{fpasym}), just using
\begin{align}
p_{gauss} (\phi,t) = \frac{1}{\sqrt{4 \pi D_{\rm eff} t}} \exp \left[ - \frac{\phi^2}{4 D_{\rm eff} t} \right],    \label{pgaus}
\end{align}
(see Fig. \ref{fig4}{\bf C}). Previous authors have used asymptotic methods for computing the associated effective diffusion coefficient $D_{\rm eff}$ inherent in the formula Eq.~(\ref{fpasym}) \citep{lifson62,lindner01}. The long-standing result is \citep{lifson62}
\begin{align*}
D_{\rm eff} = \frac{\sigma^2/2}{\int_0^{2 \pi / n} \int_0^{2 \pi /n} \e^{2 h [ \cos(n\phi) - \cos (n\psi) ]/(n \sigma^2 )} \d \psi \d \phi}, 
\end{align*}
and we can compute the integrals in the denominator to find
\begin{align}
D_{\rm eff} = \frac{\sigma^2}{2 I_0 \left( \frac{\displaystyle 2h}{\displaystyle n \sigma^2} \right)}, \label{effdc}
\end{align}
where $I_0(x)$ is the modified Bessel function of the zeroth kind. Eq.~(\ref{effdc}) demonstrates the monotone increasing dependence of the effective diffusion upon the number of potential wells $n$. %The Gaussian (\ref{fpasym}), with $D_{\rm eff}$ as a function of $n$, is plotted in Fig. \ref{fig4}{\bf C}, comparing it with the numerically calculated probability density computed from realizations of the stochastic process (\ref{sinsde}). The variance of the full stochastic process (\ref{sinsde}) scales roughly linearly with a constant approximated by (\ref{effdc}) (Fig. \ref{fig4}{\bf D}). This asymptotic calculation works well, even for small $n$ (Fig. \ref{fig4}{\bf E}). \\

To calculate the probability density $p(\phi,t)$, we simulated 10000 realizations of Eq.~(\ref{sinsde}) using Euler-Maruyama with a timestep {\tt dt = 0.001} from $t=0$ to $t=10$s (Fig. \ref{fig4}{\bf B,C}). The effective diffusion coefficient was calculated as the gradient of the variance across the time window and converted to degrees with the change of variables $\theta = 180 (\phi+ \pi)/ \pi$ so
\begin{align}
D_{\rm eff}( \theta) = \frac{180^2 D_{\rm eff}(\phi)}{\pi^2} = \left[ \frac{180^2}{\pi^2} \right]  \left[ \frac{\sigma^2}{2 I_0 \left( \frac{\displaystyle 2h}{\displaystyle n \sigma^2} \right)} \right] .  \label{deffdeg}
\end{align} \\

{\em Adding unstructured heterogeneity.} Effects of unstructured heterogeneity were studied by perturbing the potential in Eq.~(\ref{sinsde}) with a combination of random periodic functions, so
\begin{align}
\d \phi (t) = \left( - h \sin \left[ n \phi (t) \right] - \eta \frac{\d U_p( \phi)}{\d \phi} \right) \d t + \sigma \d W (t).  \label{unhetrad}
\end{align}
The unstructured heterogeneity was given by the random potential
\begin{align}
U_p( \phi )  = \sum_{j=1}^{N_h} \left[ a_j \cos (j \phi ) + b_j \sin ( j \phi ) \right],
\end{align}
where $a_j,b_j$ ($j=1,...,N_h$) are randomly drawn from normal distributions. Here $\eta$ scales the amplitude of the random potential and the maximal frequency of the unstructured components is given by $N_h$.  We take the rounded integer $N_h = [0.05(1 + 0.1\xi)/(\eta + 0.0005)+1]$, $\xi$ is a normally distributed random variable, so larger $\eta$ values reduce the number of modes added to the potential, decreasing the maximal number of attractors in the system, as in other studies of unstructured heterogeneity in bump attractor networks \citep{zhang96,renart03,itskov11,hansel13}. To calculate an effective diffusion coefficient $D_h$, we initialized 10000 simulations of Eq.~(\ref{unhetrad}) at $\phi (0) = 0$ and computed $D_h = \langle \phi(T)^2 \rangle / T$ for $T = 10$s. \\

\noindent
{\em Information measures.} To measure the performance of the network on a working memory task we used a Shannon measure of mutual information for a noisy channel \citep{cover06}.  We considered a channel receiving one of $m$ possible stimuli ($X$), storing an input as one of $n \le m$ possible states ($Y(t)$), and reading out the remembered stimulus as one of the original $m$ possible values ($Z$). The stored variable $Y(t)$ evolves during storage time $t \in [0,T]$ due to degradation of the initially loaded signal $Y(0)$ by dynamic noise. %The number of attractors $n$ is set by the frequency of spatial heterogeneity in both the spiking network and the particle model.
The stimuli were presented with equal probability $p_j = 1/m$ ($j=1,...,m$), so that the stimulus entropy was
\begin{align*}
H(X) = - \sum_{j=1}^m p_j \log_2 p_j = \log_2 m. %- \sum_{j=1}^m \frac{1}{m} \log_2 \frac{1}{m} = \frac{1}{m} \sum_{j=1}^m \log_2 m = \log_2 m.  \label{entropx}
\end{align*}
%There are two phases of information degradation which we refer to as {\em quantization} and {\em diffusion} error. 
The network represented a stimulus as the bump position at one of the system's $n$ attractors.  If $m$ was a multiple of $n$ ($m=qn$ with $q$ an integer) then the mapping from stimulus to loaded representation was straightforward with $Y(0)=\textrm{ceil}(X/q)$.
When $m$ was not a multiple of $n$, we allowed the potential well structure of the system to guide the loaded state to the nearest attractor. This lead to slightly non-uniform distributions of loaded stimuli. However, effects of diffusion made this slight non-uniformity insignificant, especially as the length of storage time $T$ was increased. In our theoretical calculations, we assumed that the loading algorithm maximized the entropy of the neural representation $Y(0)$; this sometimes involved random assignments from $X$ to $Y(0)$. In fact, we found our numerical results did not stray too far from this approximation (see Figs. \ref{fig7} and \ref{fig8}). %{\color{red} ZK: Do we need to justify why it is reasonable to assume the system maximizes entropy?} 

If $n = m$, then $Y(0)=X$ and the loaded representation had the same entropy as the stimulus $H(Y(0)) = H(X)$. If $n < m$ then the representation entropy was smaller than the stimulus entropy $H(Y(0))<H(X)$, since the space into which the stimulus is represented is smaller than the original stimulus space. Since the stimulus $X$ has a uniform distribution, then so does $Y$, with probability that attractor $j$ was loaded was $p_j = 1/n$ ($j=1,...,n$) and the entropy of the representation was $H(Y(t)) = \log_2 n$ (this holds for all $t$).
%\begin{align}
%H(Y) = - \hspace{-0.25cm} \sum_{j=1}^{\min(n,m)} p_j \log_2 p_j = \log_2 \left[ \min(n,m) \right].  \label{entropy}
%\end{align}
The readout of the neural representation required an expansion from $Y(T)$ to $Z$.  If $m$ was a multiple of $n$ then the expansion was straightforward with $Z$ having probability $1/q$ for $Z=q(Y(T)-1)+1, \ldots, qY(T)$ and zero otherwise.  In this case $H(Z)=\log_2m$.  If $m$ was not a multiple of $q$, we subdivided the domain into $m$ evenly spaced subdomains and assigned $Z$ accordingly, and for theoretical calculations we again assumed $H(Z) = \log_2m$. 

While $H(Y) \le H(Z)$, $H(Y)$ nevertheless set an upper limit for the mutual information between the stimulus $X$ and readout $Z$,
\begin{align}
I(X;Z) = I(Y;Z) = H(Y) - H(Y|Z).  \label{mutinf}
\end{align}
To compute the conditional entropy $H(Y|Z)$ we first calculated the probability of transition between one state and another during the diffusion phase ($0 \le t \le T$).  A direct estimate of the transition probabilities was obtained by numerically simulating many realizations of the model and estimating $p(\phi(0) \vert Z)$ where $\phi(0)$ is the center of mass of the bump at time $t=T$. We subdivided it into $n$ subdomains of equal width, and the area of each subdomain is $p_{j \to k}$, the transition probability from the loaded state $X=j$ to another state ($k \neq j$) or itself ($k=j$), where $k=1,...n$. Due to discrete translation symmetry in both systems, we expected $p_{j \to k} = p_{j+l \to k + l}$. The conditional entropy can then be computed \citep{cover06}
\begin{align}
H(Y|Z=j) = - \sum_{k=1}^n p_{j \to k} \log_2 p_{j \to k}.  \label{condent}
\end{align}

Our second method of computing the conditional entropy employed the effective diffusion coefficient $D_{\rm eff}$ associated with the probability density for locations of the bump or particle. $D_{\rm eff}$ depends upon $n$, ultimately introducing this further $n$ dependence into the mutual information Eq.~(\ref{mutinf}). Using the associated pure Gaussian probability Eq.~(\ref{pgaus}), we computed transition probabilities analytically for the case $j=1$, so
\begin{align}
p_{1 \to k} = \frac{1}{\sqrt{4 \pi D_{\rm eff} T}} \int_{a(k)}^{a(k)+ 2 \pi / n} \exp \left[ - \frac{x^2}{4 D_{\rm eff} T} \right] \d x = \frac{1}{2} \left[ {\rm erf} \frac{a(k) + 2\pi /n}{\sqrt{4 D_{\rm eff} T}} - {\rm erf} \frac{a(k)}{\sqrt{4 D_{\rm eff} T}} \right], \label{pjkanal}
\end{align}
for a set delay time $T$, where $a(k) = (2 \pi k -1)/n $ is the lower boundary of the $k$th subdomain and $k=1,...,\lceil n/2 \rceil$. Due to reflection symmetry of the Gaussian, we expected $p_{1 \to k} = p_{1 \to n-k}$. For even $n$, the ($n/2+1$)th subdomain is $(- \pi, \pi/n - \pi) \cup (\pi - \pi/n,\pi)$, which would lead to two integrals in the formula Eq.~(\ref{pjkanal}). As mentioned, the transition probabilities $p_{j \to k}$ for $j=2,...,n$ were easily computed as $p_{j \to k} = p_{1 \to k-j+1}$ (see Fig. \ref{fig6}{\bf B,C} insets). We then plugged each $p_{j \to k}$ value into our formula for conditional entropy Eq.~(\ref{condent}). The analytic and numeric calculation of $p_{j \to k}$ led to similar results for $I(X;Z)$, the calculated value of mutual information Eq.~(\ref{mutinf}) (Fig. \ref{fig7}).

%We then identified the $n=n_{\rm max}$ which maximized the mutual information for fixed number of possible inputs $m$ and delay time $T$, given
%\begin{align*}
%n_{\rm max} = \arg \max_{n \in \Z^+} \left[ I(X;Z) \right].
%\end{align*}
%Notice $n$ varies over all of the positive integers $\Z^+$, but we always found $n_{max}$ lay in the domain $[1,m]$ (see Fig. \ref{fig8}).

\section{Results}

%We demonstrate how spatial heterogeneity in the structure of bump attractor networks can improve the coding of spatially localized cues.  With this modification, bumps are attracted to a discrete set of spatial locations. To analyze the effect of network architecture on the encoding process, we use of a simplified potential well model, which approximates the motion of the bump. We can quantify how quickly the bump moves away from its initial position by directly calculating an effective diffusion coefficient for the bump in terms of the spatial heterogeneity of coupling. Information theory, which measures the correspondence between the recalled and actual position of the cue, uncovers a tradeoff between the quantization error in encoding due to state discretization and the error due to diffusion during the storage period.  We also demonstrate that this tradeoff is revealed by directly calculating metrics of task performance, such as the proportion of correct trials and magnitude error in recall.

\subsection{Diffusion of bumps in a spatially homogeneous network}
  
\begin{figure}
\begin{center} \includegraphics[width=16cm]{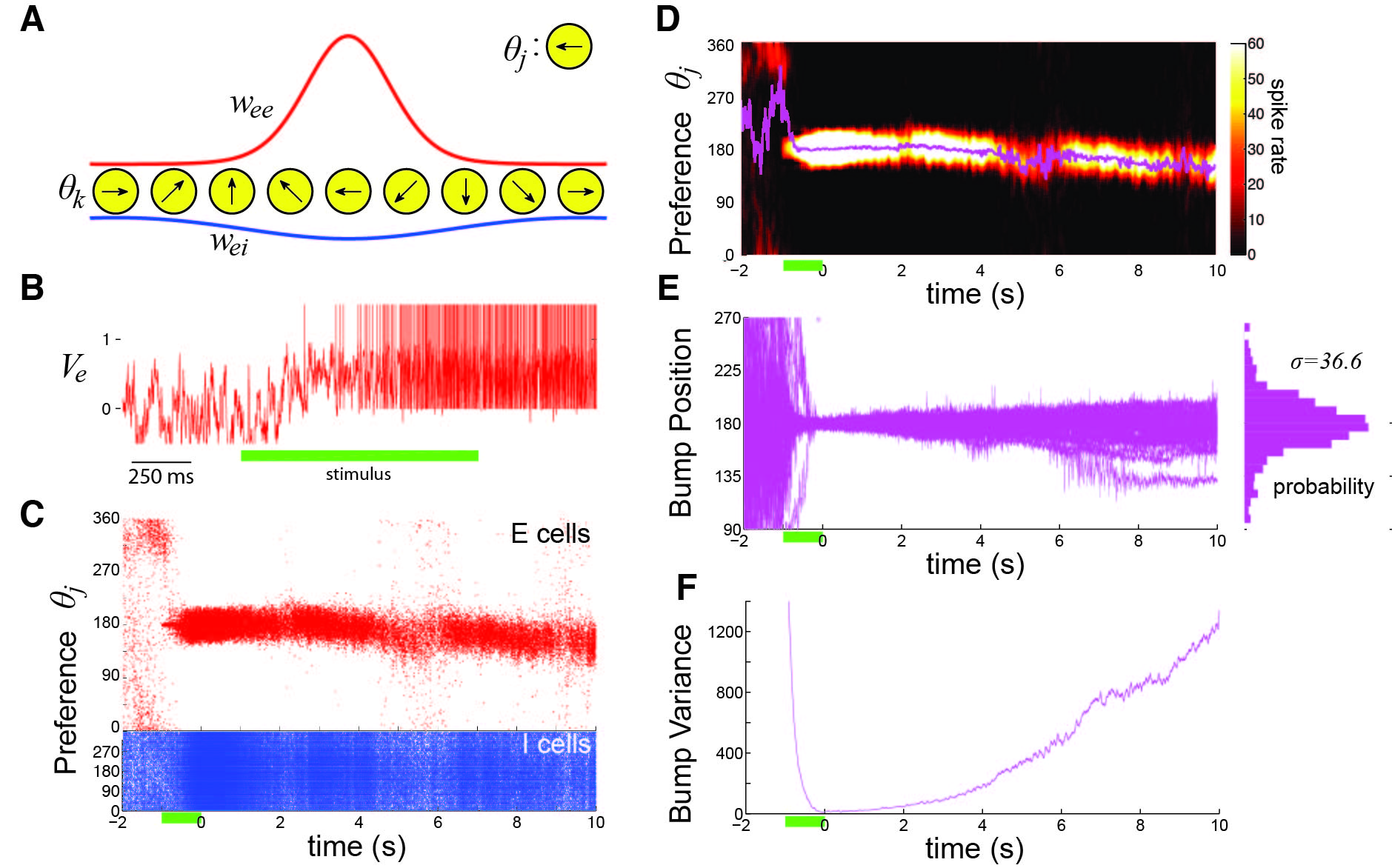} \end{center}
\caption{Spiking network model with spatially homogeneous synaptic connectivity. {\bf \em A,} Strength of connections from pyramidal to pyramidal neurons (red) and synapses from interneurons to pyramidal neurons (blue). Neuron of preferred stimulus angle $\theta_j$ receives the synaptic inputs from all neurons spanning preferred stimulus angles indexed by $\theta_k$ (see Methods). {\bf \em B,} Voltage of the pyramidal cell with stimulus preference $170^{\circ}$ before, during (green bar), and after the cue. {\bf \em C,} Formation of a bump of spiking activity in the pyramidal neurons (red) following cue presentation (green bar). {\bf \em D,} Spike rate, locally averaged across space and time (see Methods), plot shows the position of the bump's peak ($\Delta (t)$: magenta) diffuses in space, due to voltage and synapse noise. {\bf \em E,} Bump position ($\Delta (t)$) plotted for 32 realizations. The resulting probability density of bump positions from 1000 realizations after 10 seconds is roughly Gaussian. {\bf \em F,} Variance of the bump's position ($\langle \Delta (t)^2 \rangle$), across 1000 realizations, scales roughly linearly as a function of time.}
\label{fig1}
\end{figure}

The neural mechanics of parametric working memory has a long history of theoretical investigation \citep{amari77,camperi98,compte00,wang01,laing01,renart03,brody03}.  Motivated by the working memory of visual cue orientations we consider a network of spiking model neurons where each neuron has a preferred orientation in its feedforward input.  Persistent neural spiking within the network is due to a combination of assumptions on synaptic connectivity \citep{goldmanrakic95,rao99,lewis00}. First, the strength of pyramidal to pyramidal connectivity decreases as the distance between the tuning peak of each neuron increases (Fig. \ref{fig1}{\bf A}, red line). Second, excitatory synaptic currents involve both fast-acting AMPA and slow-acting NMDA components (see Materials and Methods). Third, feedback connections from interneurons are broadly tuned (Fig. \ref{fig1}{\bf A}, blue line). %We will challenge the first assumption in subsequent results, but for now we will show the effect of all synaptic connectivity being purely distance-dependent. 
With these architectural features neurons in the network respond to a transient stimulus (green bar in Fig. \ref{fig1}) with an elevated rate of spiking that persists long after the stimulus ceases (Fig. \ref{fig1}{\bf B}).      
Short-range excitation leads to high rate pyramidal spiking across a short range of orientations, while wide-range inhibition localizes this spiking (Fig. \ref{fig1}{\bf C}); we refer to this pattern of activity as a ``bump." 
%Prior to the cue, spiking is not locally focused in this way (Fig. \ref{fig1}C).
The position of the bump encodes the initial stimulus position in working memory \citep{compte00,wang01,brody03}. 

We model the inherent trial-to-trial variability of neural response with an orientation independent fluctuating input to each neuron, as well as a stochastic component of the recurrent synaptic feedback (see Materials and Methods).  These fluctuations degrade the storage of the orientation cue by causing the bump to stochastically wander away from its initial position (Fig. \ref{fig1}{\bf C,D}). Spatio-temporal averaging of the spike time raster plots identifies the maximal firing rate at each time point, and visualizes the bump wandering across the network (Fig. \ref{fig1}{\bf D}, magenta line).  We fix the stimulus orientation and perform many trials of the network simulation, with the only difference between trials being the realization of the stochastic forces in the network. The bump's position after a delay period of 10s can be described by a probability density having an overall Gaussian profile (Fig. \ref{fig1}{\bf E}), and the variance in bump position increases linearly as a function of time (Fig. \ref{fig1}{\bf F}).  These last two properties suggest the bump position behaves as a diffusion process \citep{risken96}.

Diffusive dynamics in working memory networks have been studied in several different frameworks \citep{compte00,miller06,wu08,polk12,burak12,kilpatrick13}.  The intuition for the diffusive character of these networks is best gained from an analysis of the deterministic network.
% In the absence of noise the network has neutrally stable bump solutions along orientation space.  This means that
A bump can be formed with its center of mass located at any orientation, allowing for the storage of a continuum of stimuli \citep{amari77,camperi98}. However,
%while bump states are stable to perturbations that change their size,
%the neutral stability along orientation space does not confer stability to perturbations that change bump position.
perturbations that change the bump's position will be integrated and stored as if they were another input.  Stochastic inputs lead to a continuous and random displacement of the bump, without the bump relaxing back to its original location. Over time, the position of the bump effectively obeys Brownian motion and recall error increases with the delay period. This diffusion based error is consistent with psychophysical studies which show the spread of recalled continuous variables scales sublinearly with time \citep{white94,ploner98}.   

\subsection{Reduced diffusion in a spatially heterogeneous network}

\begin{figure}
\begin{center} \includegraphics[width=16cm]{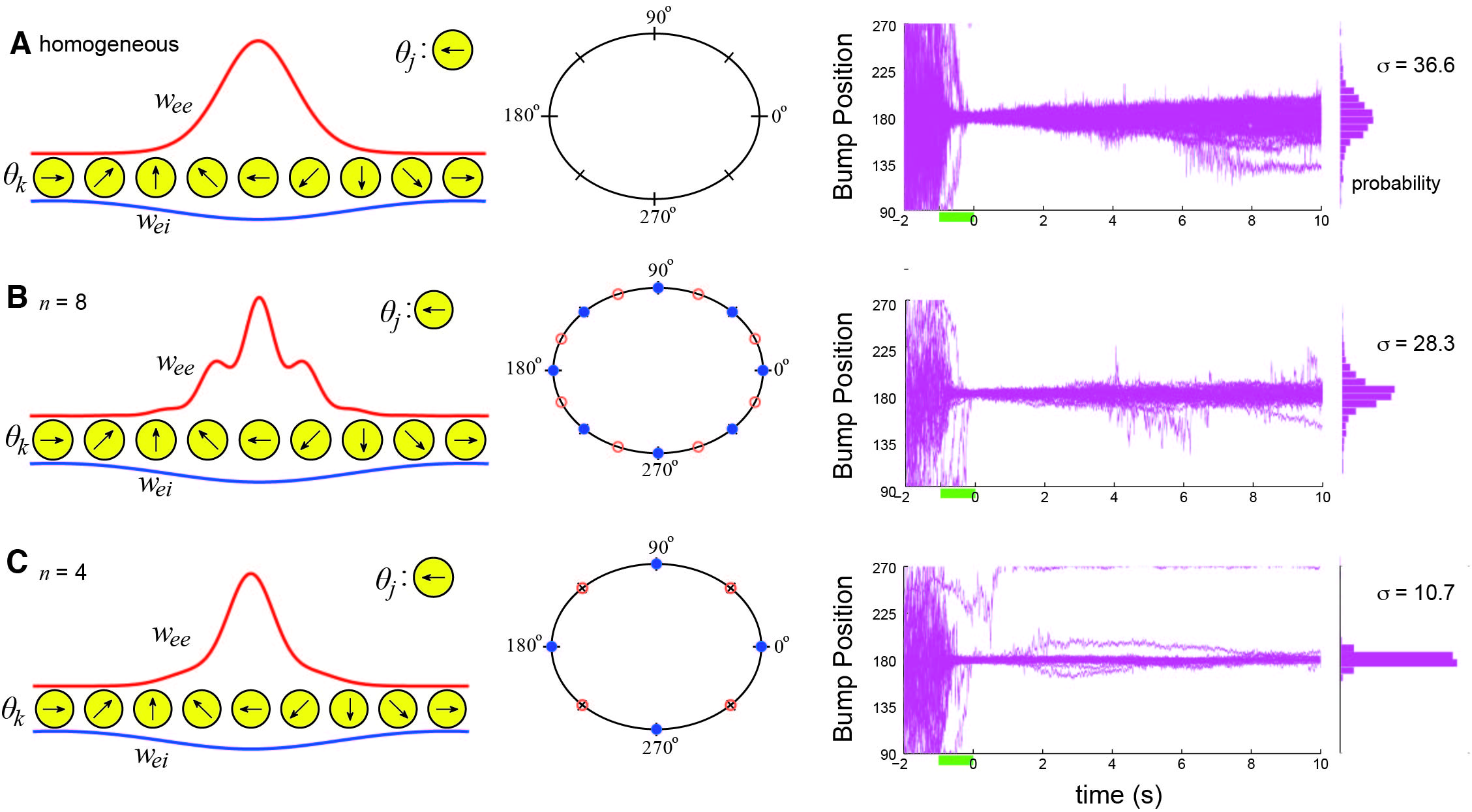} \end{center}
\caption{{\bf \em A,} Purely distance-dependent synaptic connections lead to a spatially homogeneous system. Bump dynamics lie on a line attractor, so bumps diffuse with ease {\bf \em B,} Periodically breaking the spatial homogeneity of synaptic connections with $n=8$-fold heterogeneity leads to bump dynamics evolving on a chain of $n=8$ attractors or nodes each separated by repelling states or saddles (blue). Bumps do not wander away from their initial position as easily (position plot), which tightens the resulting probability density after 10 seconds. {\bf \em C,} Effect of synaptic heterogeneity is more noticeable for a $n=4$-fold break in homogeneity. Bump position rarely strays from $180^{\circ}$ as shown by the very tight probability density.}
\label{fig2}
\end{figure}

\begin{figure}
\begin{center} \includegraphics[width=10cm]{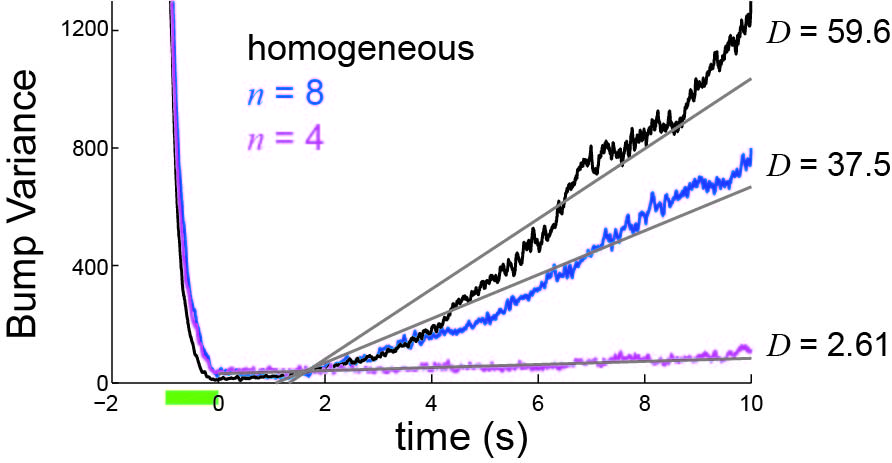} \end{center}
\caption{Variance of bump position as a function of time, averaged across 1000 realizations for spatially heterogeneous structure of pyramidal to pyramidal synapses with frequency $n=4$ and $n=8$ as well as spatially homogenous structure. We fit each curve to straight lines to generate an approximation of the effective diffusion coefficient $D$ (see Methods).}
\label{fig3}
\end{figure}

Previous models of working memory have considered networks that use neuronal units with bistable properties \citep{rosen72,koulakov02,goldman03,brody03,miller06}.  These networks lack the homogeneity required for a continuum of neutrally stable stimulus representations, and rather have a discrete number of stable states.  One advantage of this network heterogeneity is a `robustness' of representation with respect to parameter perturbation, a feature that is absent in homogeneous networks \citep{goldman03,brody03}.  We consider spatially periodic modulation of excitatory coupling (Fig. \ref{fig2}, left), where the period of the modulation is $360/n$ degrees, so that $n$ cycles cover orientation space.  We assume such an architecture would not be biased to favor one particular cue location because errors reported in recalling cues are roughly the same for each cue location \citep{white94}. Such an architecture may develop from Hebbian plasticity rules during training in working memory tasks, since orientation cues are typically chosen at fixed and evenly spaced locations around the circle \citep{funahashi89,white94,goldmanrakic95,meyer11}. In this situation some neuron pairs are activated more than other pairs, leading to relative strengthening of their recurrent connections \citep{clopath10,ko11}. Alternatively, reward-based plasticity mechanisms could also set up spatial heterogeneity in synapses if it improved a subject's performance during a task \citep{schultz98,wang08,klingberg10}.   

Spatial biases introduced to network architecture shift the smooth continuum of stable states to a chain of discrete attractors, each separated by a repeller (compare Fig. \ref{fig2}{\bf A} to Figs. \ref{fig2}{\bf B,C}, middle column). This discrete attractor structure occurs because some pyramidal neurons receive stronger excitatory projections than others \citep{zhang96,itskov11,hansel13}. Spatial heterogeneity in the strength of excitatory connections (decreasing $n$) stabilizes bump positions to perturbations by noise (compare Fig. \ref{fig2}{\bf A} to Figs. \ref{fig2}{\bf B,C}, right column). For all $n$ tested, the probability density of bump positions retains an approximately Gaussian shape with periodic modulation, so the variance of bump positions still grows nearly linearly with time (Fig. \ref{fig3}), and it is well approximated by $Dt$ where $D$ is the diffusion coefficient (see Materials and Methods).  The coefficient $D$ drops considerably for a network with spatially heterogeneous synapses, compared to the bump diffusion measured with homogenous network.  In total, spatial heterogeneity of excitatory coupling helps stabilize bump position in models of working memory with fluctuating stochastic inputs.

\subsection{Potential well model for bump diffusion}

\begin{figure}
\begin{center} \includegraphics[width=15cm]{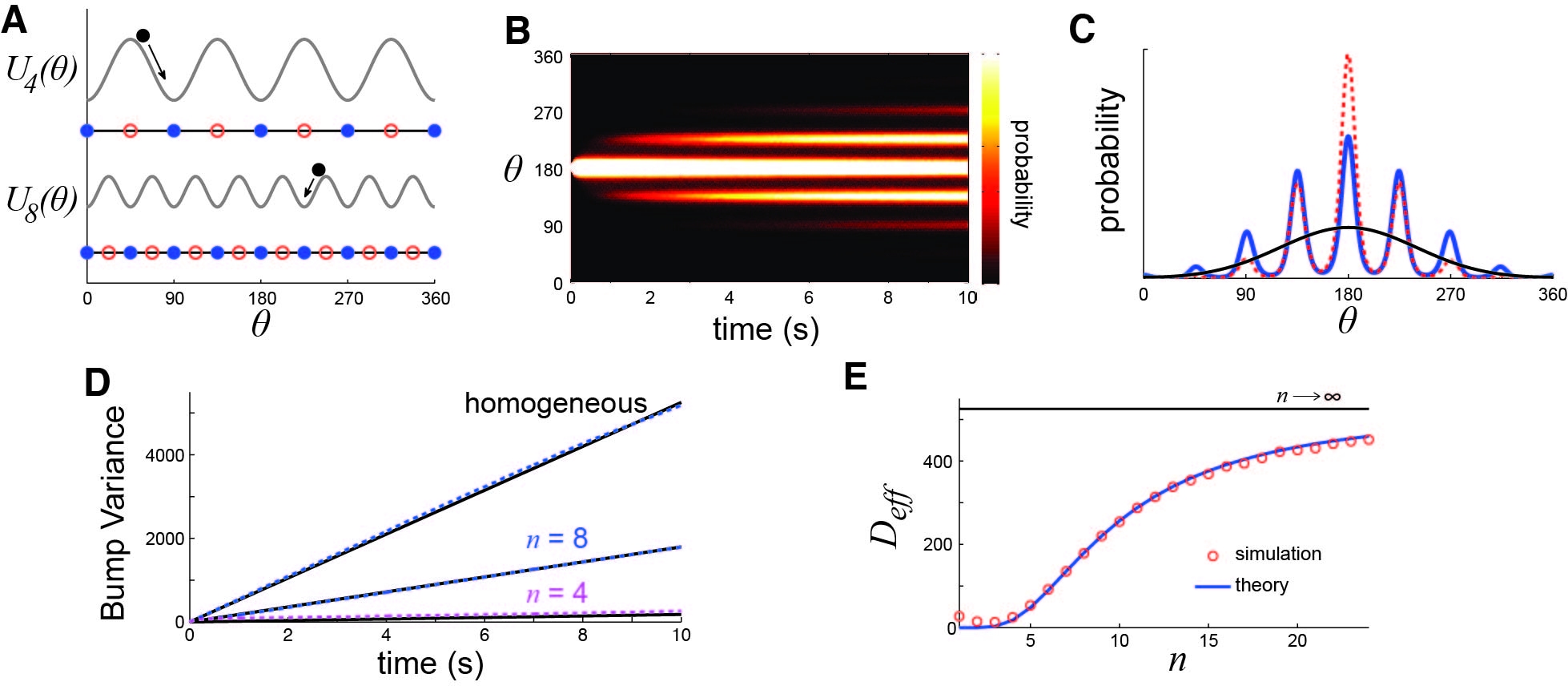} \end{center}
\caption{{\bf \em A,} Particle diffusing in periodic potential well Eq.~(\ref{cospot}) is an idealized model of the bump diffusing the network with spatially heterogeneous synapses, given by stochastic process Eq.~(\ref{sinsde}). {\bf \em B,} Probability density $p(\theta,t)$ of particle position $\theta$ spreads diffusively in time ($n=8$; $h=1$; noise variance $\sigma^2 = 0.16$). {\bf \em C,} Profile of $p(\theta,t)$ computed from 10000 realizations (red); effective diffusion theory Eq.~(\ref{fpasym}) (black); and effective diffusion theory with periodic correction Eq.~(\ref{pgaus}) (blue). {\bf \em D,} Effective diffusion theory using $D_{\rm eff}$ Eq.~(\ref{effdc}) matches variance scaling from simulations of stochastic equation very well. Variance increases monotonically with the well frequency $n$. {\bf \em E,} An effective diffusion coefficient $D_{\rm eff}$ can be computed by treating well hopping as a jump Markov process \citep{lifson62,lindner01}, yielding formula Eq.~(\ref{effdc}).}
\label{fig4}
\end{figure}

To analyze the relationship between network heterogeneity and bump diffusion more deeply, we now study an idealized model for parametric working memory. Briefly, a noise driven particle on a periodic potential landscape retains the essential effects of noise and spatial heterogeneity in our spiking network model (see Materials and Methods). Our simplified model treats the bump position as a particle moving in a landscape of peaks, from which it is repelled (maxima in Fig. \ref{fig4}{\bf A}), and wells, to which it is attracted (minima in Fig. \ref{fig4}{\bf A}). %Occasionally, noise will propel the particle over a hill to the next valley. 
In the potential well model the memory of the stimulus location is tracked by the particle's position $\theta(t)$ obeying the following stochastic differential equation: 
\begin{align}
\d \theta (t) = \left( - h \sin \left[  \frac{180}{\pi} n \theta (t) \right] \right) \d t + \d W(t).  \label{thlang}
\end{align}
Here $h$ is the amplitude of the periodic potential and $W(t)$ is a Wiener process \citep{risken96}.
The sine function determines the $\theta$-dependent drift of the particle, which ultimately affects its diffusive motion. The positive integer $n$ determines the number of stable attractors.  Similar reduced neural models have been explored \citep{renart03,itskov11}, and the general problem of noise-induced behavior in periodic potentials has been well studied \citep{lifson62,risken96,lindner01}.  

Diffusive behavior occurs in periodic potentials, yet the mechanics is different than diffusion on a free potential landscape ($h=0$).   In periodic potentials, a particle typically undergoes small variance dynamics confined within a well.  However, rare but large noise kicks eventually push the particle to a neighboring well.  These noise-induced well transitions continue indefinitely, and diffusion over the potential landscape occurs in a punctuated fashion. Across many trials, the probability $p(\theta, t)$ of finding the particle at position $\theta$ at time $t$ evolves like a Gaussian kernel modulated by a periodic function (Fig. \ref{fig4}{\bf B}, see Materials and Methods).  The maxima of $p( \theta, t)$ are centered at the minima of the potential, indicating a higher likelihood of finding the particle at the bottom of a well than in transition between wells.  

Treating the transitions between wells as a jump process, we can approximate the diffusion of the sine potential model, Eq.~(\ref{thlang}) with the following Brownian walk: 
\begin{align}
\d \theta (t) = \sqrt{2 D_{\rm eff}(n)} \d W (t). 
\end{align}
The effective diffusion coefficient $D_{\rm eff}(n)$ is derived using standard approaches (see Eq. (\ref{deffdeg}) in the Materials and Methods).  Under this approximation, $\theta(t)$ obeys a Gaussian distribution with variance $D_{\rm eff}(n)t$, and $p(\theta,t)$ does not possess the periodic microstructure of the actual probability density (Fig. \ref{fig4}{\bf C}, compare blue and black curves).  %However we can enhance the model by multiplying the Gaussian kernel by the stationary density $p_0( \phi )$ (Fig. \ref{fig4}{\bf C}, compare blue and dash-red curves; see Materials and Methods).  
Despite these differences, the approximation agrees very accurately with the variance in the particle's position in the periodic potential (Fig. \ref{fig4}{\bf D}).  Thus, in this framework we can directly relate the frequency of heterogeneity $n$ to the overall diffusivity of the particle through $D_{\rm eff}(n)$. As with bumps in the spiking network, increasing the frequency $n$ of the spatial heterogeneity increases the effective diffusion $D_{\rm eff}$ (Fig. \ref{fig4}{\bf D}). This is true across the entire range of frequencies $n$, and as $n \to \infty$ the particle's variance saturates to that of a system with a flat potential (Fig. \ref{fig4}{\bf E}).  Thus, despite the simplicity of the potential well framework it can qualitatively explain the diffusivity observed in the spiking network. In both Fig. \ref{fig3} and \ref{fig4}{\bf D}, the variance in bump position is fit well by a linear function whose slope decreases with the number of attractors $n$. 

\subsection{Impact of unstructured heterogeneity}

As we have shown, a structured heterogeneity of cortical architecture that generates evenly spaced attractors (Figs. \ref{fig2},\ref{fig4}A) curtails the rate of diffusion (Figs. \ref{fig3},\ref{fig4}{\bf D}). However, synaptic architecture may also have random components that are neither related to task specifics nor optimized to any specific computation \citep{wang06}.  In principle, such perturbations in architecture could degrade the performance of working memory networks that require a fine tuned architecture.   \cite{renart03} demonstrated that the deleterious effects of such unstructured heterogeneity in bump attractor networks can be mitigated by homeostatic plasticity, which spatially homogenizes network excitability. In our model, considering such a process would bar the system from establishing spatially structured heterogeneity, which we have shown improves storage accuracy. Thus, we next study how a combination of structured and unstructured spatial heterogeneity affects the diffusive dynamics in working memory networks. We show that the system is still robust to noise, even when the potential is altered in this way.

\begin{figure}
\begin{center} \includegraphics[width=11.5cm]{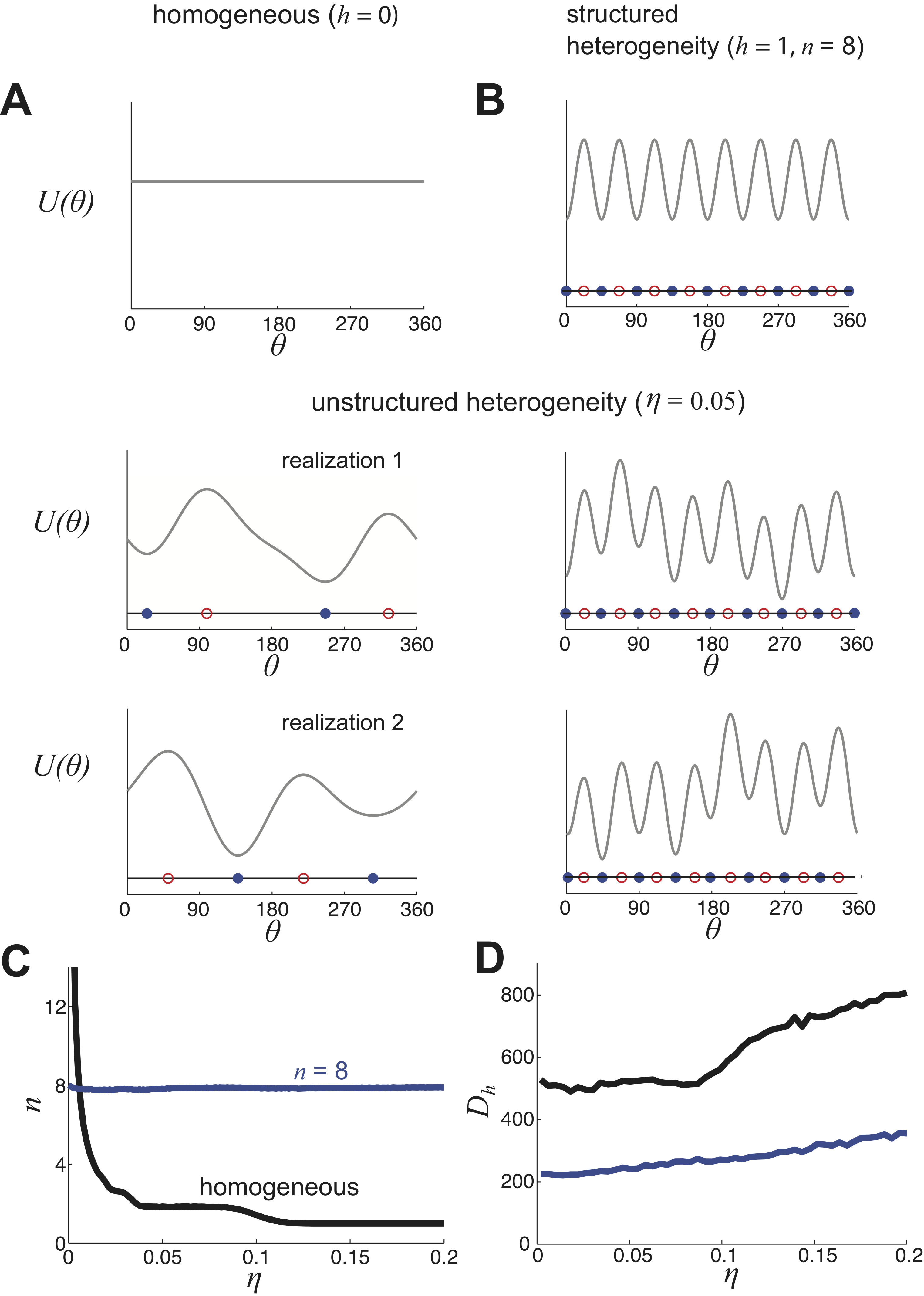} \end{center}
\caption{Potentials with structured heterogeneity resist degradation from unstructured heterogeneity. {\bf \em A,} Adding unstructured heterogeneity to a flat potential function drastically alters the state space of attractors. {\bf \em B,} When structured heterogeneity is already present, adding unstructured heterogeneity does not change the number of attractors or their positions. {\bf \em C,} The number of attractors $n$ is strongly influenced by the severity of unstructured heterogeneity $\eta$ in the homogeneous potential. Starting with $n=8$ attractors, adding unstructured heterogeneity does not alter $n$. {\bf \em D,} Effective diffusion increases much more for the homogeneous potential as a function of $\eta$ than for the potential containing structured heterogeneity.}
\label{fig5}
\end{figure}

Specifically, we modify the shape of the potential in Eq.~(\ref{thlang}), so that
\begin{align}
\d \theta (t) = \left( - h \sin \left[ \frac{180}{\pi} n \theta (t) \right] - \eta \frac{\d U_p (\theta)}{\d \theta} \right) \d t + \d W (t),  \label{potpert}
\end{align}
where $U_p(\theta)$ is an unstructured perturbation of the underlying cosine potential function (see Materials and Methods).  Briefly, $U_p(\theta)$ is a component of the potential that is randomized from trial-to-trial (two realizations of the full potential are shown in Fig. \ref{fig5}{\bf A,B}).    %We study the influence of $U_p(\theta)$ on bump diffusion in the presence ($h>0$) or absence ($h=0$) of structured heterogeneity. %For transparency, we demonstrate our findings in the potential well model, but we believe they extend to the spiking network as well.
Adding a small component of unstructured heterogeneity ($\eta >0$) to a network with an initially flat potential function ($h=0$) substantially alters the attractor structure (Fig. \ref{fig5}{\bf A}). The network shifts from having a continuum of preferred locations to having a small number of preferred locations at disordered positions. This is analogous to the drastic collapse in the number of possible stable bump locations observed in bump attractor models whose synaptic structure is randomly perturbed \citep{zhang96,renart03,itskov11,hansel13}. On the other hand, a network that possesses structured heterogeneity ($h>0$) retains the original positions of its stable attractors after unstructured heterogeneity is added, even though the profile of the potential is distorted (Fig. \ref{fig5}{\bf B}). When the severity of heterogeneity is increased (larger $\eta$), the number of attractors is considerably reduced in the network without structured heterogeneity, while remaining the same in the network with structured heterogeneity (Fig. \ref{fig5}{\bf C}).
Lastly, the effective diffusion $D_h$ (see Materials and Methods) of the network containing structured heterogeneity increases only gradually as the degree of unstructured heterogeneity is increased (Fig. \ref{fig5}{\bf D}), contrasting the distinct rise in the effective diffusion $D_h$ in the network without structured heterogeneity. 

Therefore, the spatial organization of attractors in the network with structured heterogeneity is robust to random perturbations of the underlying potential landscape. Recent studies have shown that parametrically-perturbed spiking network models of bumps retain dynamics whose spatial profile is bump-shaped \citep{brody03,itskov11,hansel13,kilpatrick13}. \cite{brody03} showed the effective dynamics of the resulting system can then be numerically approximated by a potential well model like Eq.~(\ref{potpert}). Thus, the low-dimensional dynamics of the spiking network model can still be described by the potential well model, so we believe the spiking network will also be robust to unstructured perturbations in its spatial architecture. This robustness allows the reduction of diffusion due to structured heterogeneity to be relatively unaffected by sources of unstructured heterogeneity that undoubtedly exist in most cortical networks \citep{wang06}.  %In the analysis that follows, we focus on how structured heterogeneity engages a tradeoff between reduced diffusion and a collapsed stimulus representation space to optimize information transfer.

\subsection{Memory storage as a noisy channel}

Structured spatial heterogeneity in recurrent excitatory coupling has two distinct influences on response fidelity in working memory networks.  First, it produces a finite set of attractors with which to store stimuli.  Second, as we have shown, heterogeneity reduces the diffusion of persistent bump states across the network.  These two influences have consequences for the overall storage performance by the network. We next characterize the working memory network as a noisy information channel \citep{cover06} and show how spatial heterogeneity of excitatory coupling mitigates a tradeoff between errors due to these two influences.  
   
\begin{figure}
\begin{center} \includegraphics[width=15.5cm]{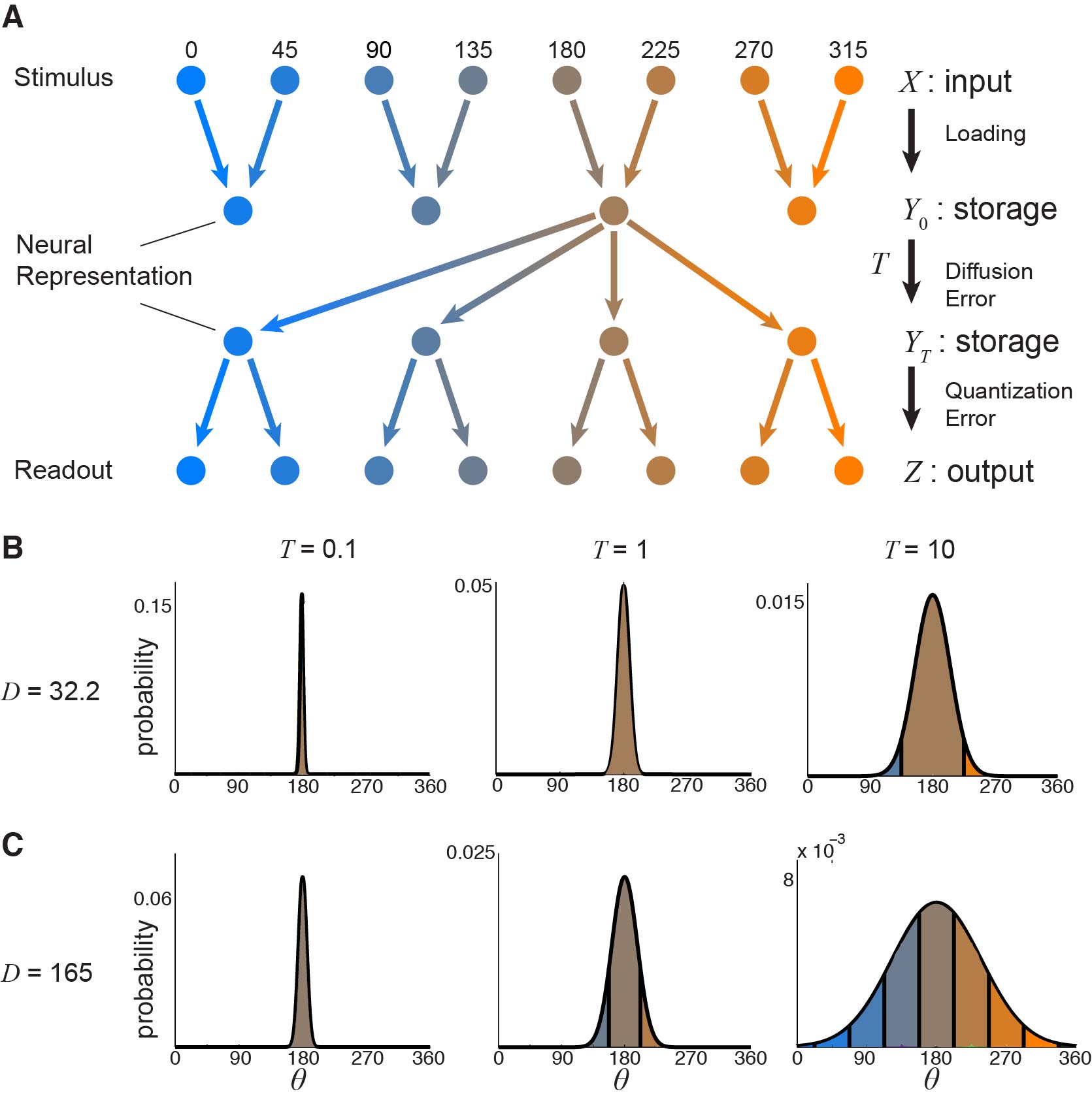} \end{center}
\caption{Noisy channel description of memory storage. {\bf \em A,} Loading $m$ possible initial conditions into $n$ possible wells initially reduces information. After the storage period ($T$), information may have been lost due to hops between wells. {\bf \em B,} Purely Gaussian probability density with the effective diffusion coefficient $D_{\rm eff}$ calculated when $n=4$ and $\sigma^2 = 0.16$. The area of each filled portion represents the probability of recalling the cue angle associated with that color. Each area corresponds to the probability of transitioning from the original state to that state $p_{j \to k}$ as visualized in the inset transition matrices. {\bf \em C,} For $n=8$, the effective diffusion coefficient $D_{\rm eff}$ is larger, leading to faster spreading.}
\label{fig6}
\end{figure}

Consider a stimulus chosen from $m$ equally likely values which is to be stored by a working memory network.  The network has $n$ attractors and must store the stimulus value for $T$ seconds before being read out.  From a coding perspective we have a chain where random input $X \in [1,m]$ is loaded into attractor $Y(0) \in [1,n]$ and remains in storage until $Y(T) \in [1,n]$, after which it is finally readout as response $Z \in [1,m]$ (Fig. \ref{fig6}{\bf A}).  If $n<m$ then the transition $X \to Z$  involves a compression ($X \to Y(0)$) and expansion ($Y(T) \to Z$) of data, causing errors in transmission due to the quantization of the neural representation (Materials and Methods).  

The transition $Y(0) \to Y(T)$ involves diffusion across the network, which also degrades storage.  To compute the probability of transitioning from one attractor to another during the storage phase, we need only integrate the Gaussian approximation with variance $D_{\rm eff}(n)T$ over the appropriate domain (Fig. \ref{fig6}{\bf B,C}). In this way, we calculate the matrix of transition probabilities from one attractor to another during the retention interval.  With this matrix we can calculate the information lost due to diffusion (Materials and Methods).  Naturally, as the product $D_{\rm eff}(n)T$ increases, diffusion becomes more prominent (Fig. \ref{fig6}{\bf B,C}), and information loss due to diffusion increases.
 
In total, an increase in $n$ has the dual effect of reducing quantization error, yet increasing diffusion error.  Thus, we predict that spatial heterogeneity (measured by $n$) causes a tradeoff between quantization and diffusion based error, and an optimal heterogeneity will maximize the overall information flow across the channel.  We explore this prediction in the next section.   

\subsection{Optimizing information flow with heterogeneous network coupling}

Delay time $T$ and the number of possible stimuli $m$ can be easily controlled in working memory experiments \citep{funahashi89}.  By fixing the protocol in working memory tasks, it has been shown animals can improve their performance through extensive training, and boost their average reward rate \citep{meyer11}. Performance improvements are likely caused by modifications to the structure of networks underlying working memory, so we presume the spatial heterogeneity of the network, parametrized by $n$, evolves internally through reward-based plasticity mechanisms  \citep{schultz98,wang08,klingberg10}. %To maximize the likelihood of successfully recalling a stimulus, dopaminergic reward-based plasticity mechanisms may alter the spatial structure of the network to effectively tune this parameter \citep{schultz98,wang08,klingberg10}.  
To measure the overall success of storage we consider the mutual information $I$ between $X$ and $Z$ for both the potential well model and the full spiking network (Materials and Methods).  Mutual information measures the reduction in uncertainty in stimulus $X$ when response $Z$ is known.  Furthermore, mutual information allows for a clean dissection of the information loss due to quantization and diffusion based errors.  

The stimulus space compression involved in $X \to Y(0)$ involves a loss of $\log_2(m/n)$ bits, a quantity that decreases with $n$.  Calculating information loss due to diffusion  ($Y(0) \to Y(T)$) requires that we compute the effective diffusion coefficient $D_{\rm eff}(n)$.  To obtain  $D_{\rm eff}(n)$ for the potential well model we use our analytical approach (see Eq. (\ref{deffdeg}) in Materials and Methods), while for the spiking model we use a numeric fit to $D_{\rm eff}(n)$ (Fig. \ref{fig3}).  Information loss due to diffusion increases with $D_{\rm_eff}(n)$, which in turn increases with $n$ (Figs \ref{fig3}, \ref{fig4}{\bf E}).  Given both sources of information loss, we compared the channel theory prediction for $I(X,Z)$ to direct estimates of $I(X,Z)$ based on the joint density $p(X,Z)$ (Materials and Methods), for both the spiking and potential well models. %Using mutual information to measure errors quantifies the spread of the probability density after a set period of time. The more spread out the density is, the greater the loss of information due to error. This provides an analog description of information loss. Such an analog description of the amount of error made in recalling a signal is pertinent. Making a small error in a visual judgement can result in a smaller penalty than a large error, during and outside of laboratory experimental paradigms \citep{todorov04}.     

\begin{figure}
\begin{center} \includegraphics[width=8cm]{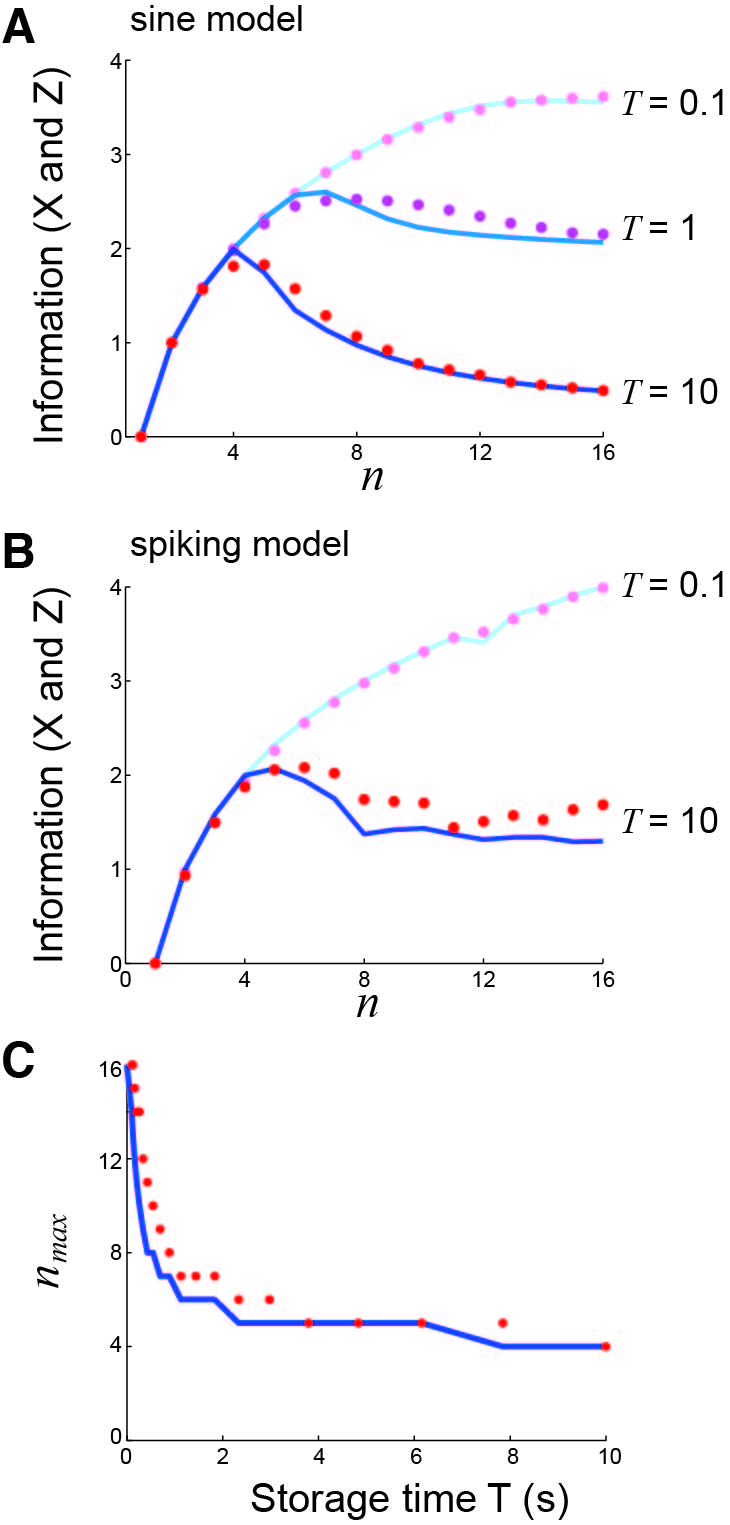} \end{center}
\caption{An optimal number of attractors $n_{max}<m$ emerges as the delay time $T$ grows with $m$ fixed. {\bf \em A,} Mutual information $I(X;Z)$ between input particle position $X$ and output recalled position $Z$ varies with $n$ in the sine model for delay (storage) times $T=0.1$, $T=1$, and $T=10$; input number $m=16$; well height $h=1$; and noise variance $\sigma^2 = 0.16$. {\bf \em B,} Mutual information $I(X;Z)$ calculated between initial $X$ and final $Z$ position of the bump in the spiking network for delay times $T=0.1$ and $T=10$ with $m=16$ inputs. {\bf \em C,} Keeping the number of possible initial positions $m=16$ fixed reveals that $n_{\rm max}$ decreases monotonically with delay time $T$. As the well height $h$ increases, these curves shift to higher values of $n_{\rm max}$.}
\label{fig7}
\end{figure}

For a fixed $m$ and very short delay time $T$ the information $I(X,Z)$ monotonically increases with $n$, and is maximized when $n=m$ (Fig. \ref{fig7}{\bf A,B} $T=0.1$s). This is expected, since recall is near immediate, so that diffusion-based error is negligible and quantization error dominates the information loss.  This error vanishes when $n=m$ and $I(X,Z)$ approaches the stimulus entropy ($\log_2(m)$ bits).  As the delay time is increased, information loss due to quantization error does not change, but errors due to diffusion increase. For sufficiently long $T$, information peaks at a value of $n_{\rm max}<m$ in both the potential well model and the spiking model (Fig. \ref{fig7}{\bf A,B} $T=10$s).  The value $n_{\rm max}$ marks a compromise between quantization and diffusion errors.  For the potential well model the optimal heterogeneity $n_{\rm max}$ decreases as the delay time $T$ increases (Fig. \ref{fig7}{\bf C}), since diffusion error grows as $T$ increases.  In total, we find that for sufficiently long delay times the degree of heterogeneity $n$ should be less than the stimulus size $m$ to optimize information transfer. 

\begin{figure}
\begin{center} \includegraphics[width=8cm]{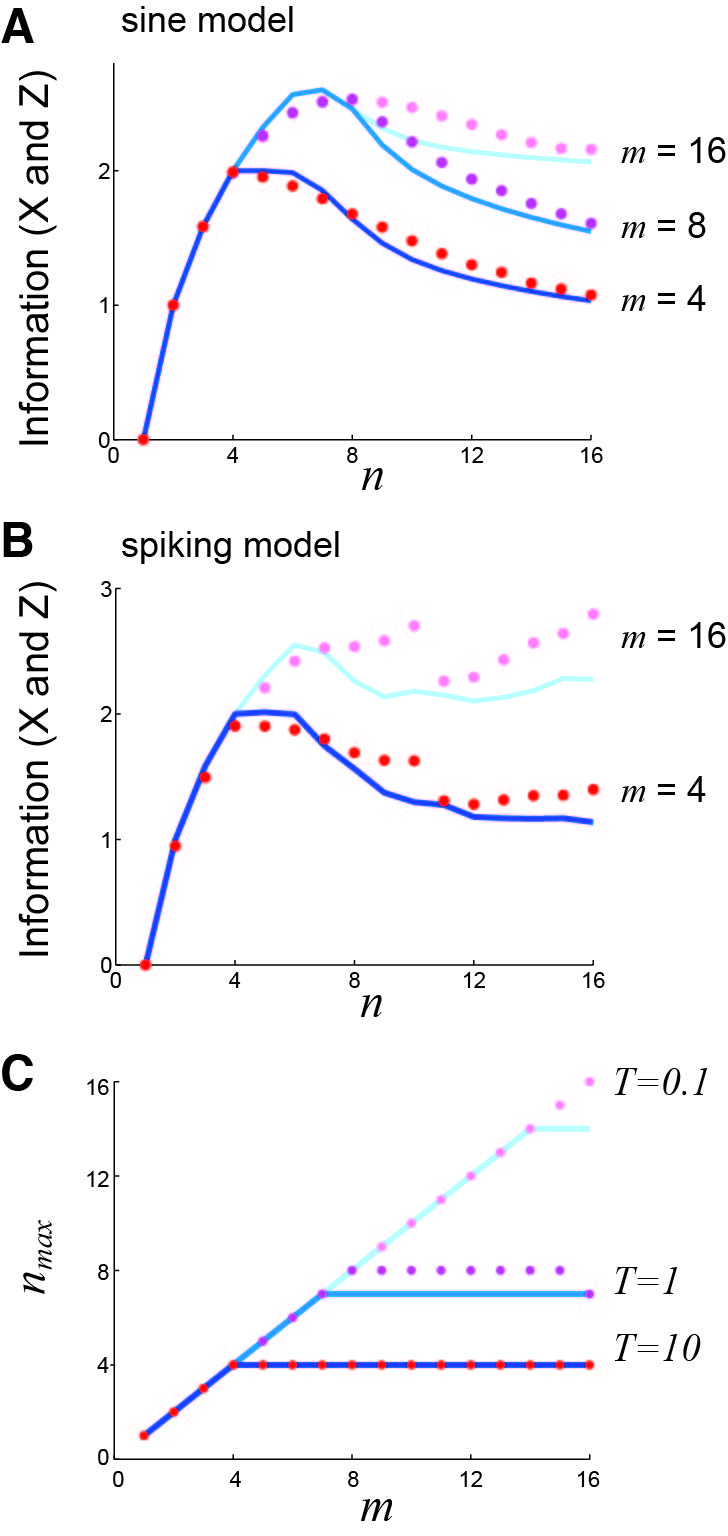} \end{center}
\caption{An optimal number of attractors $n_{max}<m$ emerges as $m$ is increased with $T$ fixed. {\bf \em A,} $I(X;Z)$ in sine model for input numbers $m=4$, $m=8$, and $m=16$ with delay time $T=1$ fixed. {\bf \em B,} $I(X;Z)$ in spiking model for input numbers $m=4$ and $m=16$ with delay time $T=5$ seconds fixed. {\bf \em C,} Fixing $h=1$ shows $n_{\rm max}$ generally increases with $m$. As the delay time $T$ increases, $n_{\rm max}$ reaches an optimum at smaller cue numbers $m$. Noise variance $\sigma^2 = 0.16$.}
\label{fig8}
\end{figure}

For a fixed delay time $T$, varying the number of possible inputs $m$ also shifts $n_{\rm max}$. Diffusion error is independent of the number of possible inputs $m$; however, the total possible information increases with the number of inputs $m$. For small $m$ we have that $n_{\rm max}=m$ since when $n\ge m$ the quantization error is always zero and network diffusion increases with $n$ (Fig. \ref{fig8}{\bf A,B}, $m=4$).  However, for larger $m$, we find $n_{\rm max}<m$, due to a compromise between quantization error and diffusion (Fig. \ref{fig8}{\bf A,B}, $m=16$).  These results hold for the potential well model over a wide range of $m$ (Fig. \ref{fig8}{\bf C}).  
Overall, we highlight that a combination of varying $T$ and $m$ uncovers the effect of diffusion and quantization error on mutual information in a working memory network. In particular, for many combinations of $T$ and $m$ an optimal spatial heterogeneity for information transfer can be found.  

\begin{figure}
\begin{center} \includegraphics[width=12cm]{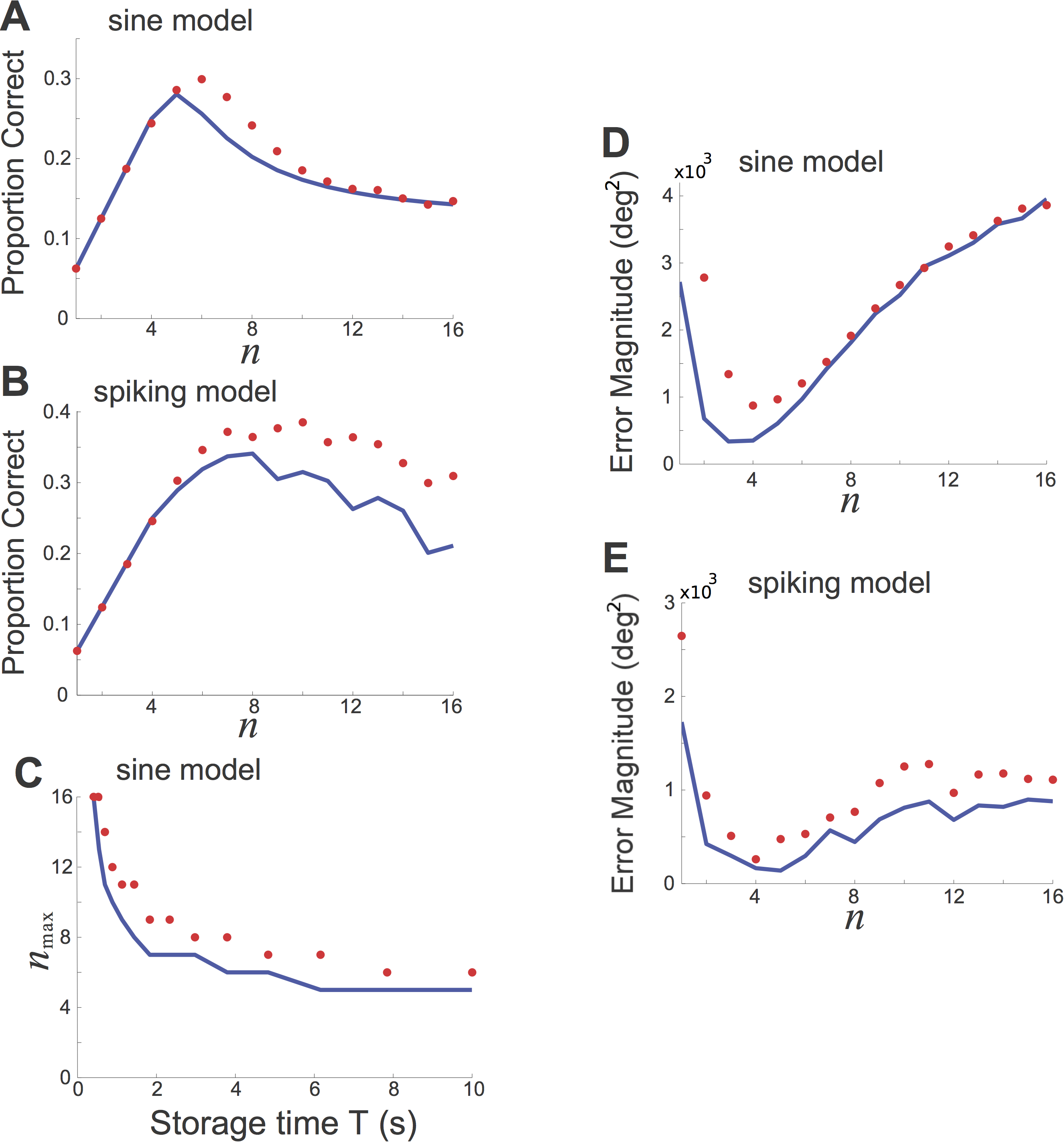} \end{center}
\caption{Proportion correct and error as it depends on the number of possible outputs, $n$. Blue curves are computed using theoretical probability densities and red dots employ numerically computed probability densities (see Methods). {\bf A}, Proportion $p(Z=X)$ of responses $Z$ same as the input $X$ varies with $n$ in the sine model for delay (storage) time $T=10$s. {\bf B}, Proportion of correct responses $p(Z=X)$ as a function of $n$ in the spiking model for delay (storage) time $T=10$s and input number $m=16$. {\bf C}, The number of outputs $n_{max}$ that maximizes the proportion of correct responses as the storage time $T$ is varied in the sine model.  {\bf D}, Error magnitude (in degrees$^2$) varies with $n$ in the sine model for delay (storage) time $T=10$s. {\bf E}, Error magnitude (in degrees$^2$) varies with $n$ in spiking model for delay (storage) time $T=10$. Input number $m=16$. In the sine model, well height $h=1$ and noise variance $\sigma^2=0.16$.}
\label{fig9}
\end{figure}

The information $I(X,Z)$ measures the general relation between $X$ and $Z$, one that is decoder independent.  However, in psychophysical experiments a reward is only given when the recall is correct, i.e $X=Z$.  Thus, it is important to consider how the probability of correct recall depends on the spatial heterogeneity of excitatory connections.  In both the potential well and spiking network models the probability of correct recall is maximized for a fixed $n < m$ when $T$ is sufficiently long (Fig. \ref{fig9}{\bf A,B}), consistent with observations of $I(X,Z)$ (Fig. \ref{fig7}{\bf A,B}).  The $n$ that maximizes the probability of correct recall decreases as storage time increases (Fig. \ref{fig9}{\bf C}), also in agreement with results using $I(X,Z)$ (Fig. \ref{fig7}{\bf C}).  The probability of correct recall provides a quantification of error that weights all incorrect responses the same.  To use knowledge of the spatial organization of the cue set in determining error, we also measure the impact of the number of attractors on the angular difference between the recalled and cued stimulus position. Specifically, we compute the variance of the difference between the recalled and input cue location $X-Z$. The magnitude of the recall error is minimized for a fixed $n<m$ (Fig. \ref{fig9}{\bf D,E}), corroborating our findings for $I(X,Z)$ and proportion correct. Thus, our core finding that information transfer across the memory network is maximized for a specific degree of spatial heterogeneity also holds for a measure of task performance.  %This last result links our work to behavioral correlates of working memory neural dynamics.            

\section{Discussion}
\label{disc}

We have outlined how both neural architecture and noisy fluctuations determine error in working memory codes. In working memory networks, the position of a bump in spiking activity encodes the memory of a stimulus, and input fluctuations cause diffusion of the bump position which degrades the memory.  Spatially heterogeneous recurrent excitation reduces the diffusion of bumps by stabilizing a discrete set of bump positions.  However, this also introduces memory quantization, limiting the capacity of information transfer.  By analyzing the information loss incurred by both error sources, we can maximize the transfer of information between the stimulus and the memory output by tuning the spatial heterogeneity of recurrent excitation. We found that the ideal heterogeneity gives a number of attractors in the network $n_{\rm max}$ which can be less than the number of possible inputs to the network $m$. 

\subsection{Robust bump dynamics through quantization}
Networks whose dynamics lie on a continuum attractor have steady state activity that can be altered by arbitrarily weak noise and input \citep{bogacz06}.  The advantage of this feature is that two stimuli with an arbitrarily fine distinction can be reliably stored and distinguished upon recall.  However, this structure requires fine-tuning of network architecture, since any parametric jitter will destroy a continuum attractor.  Previous work has shown how spatial heterogeneity in recurrent excitatory coupling quantizes the continuum attractor and stabilizes persistent network firing rates to perturbations in model parameters \citep{koulakov02,goldman03,brody03,cain12} and fluctuations \citep{fransen06}. We believe our results apply to these models and have extended this previous work in two major ways. 

First, we have shown that quantizing the state space of a spatially structured network into a finite number of attractor positions stabilizes bump position to dynamic noise. Second, we have shown that there is an optimal number of attractor positions for storing stimuli when dynamic noise is present. The optimal number can be lower than the actual quantization of possible stimuli, so that under-representing stimulus space can lead to more reliable coding.
%The visual system has a considerable quantization limit to distinguishing different spatial frequencies and orientations \citep{bradley87}. Thus, networks encoding working memory of a continuous parameter would do well to at least quantize representation space at this level.
Studies of networks encoding the memory of eye position show individual neurons exhibit bistability in their firing rates \citep{aksay03}, which motivated modeling their firing rate to input relations as quantized, staircase-shaped functions \citep{goldman03}. This provides an example of a working memory network thought to provide a discrete delineation of a continuous variable.
% As our results suggest, the optimal representation space for a working memory task should not have accuracy that will be degraded by diffusion error. 
Our results also suggest parametric working memory networks should coarsen the stored signal to guard against diffusion error.

\subsection{The advantage of spatial heterogeneity}
Past work has suggested spatial heterogeneity in working memory networks is a barrier to reliable memory storage \citep{zhang96,renart03,itskov11,hansel13}. In these studies, parameters of single neurons \citep{renart03} or synaptic architecture \citep{itskov11,hansel13} change throughout the network in a spatially aperiodic way. Substantial quantization error results since the bump drifts towards one of a finite number of attractor positions that may not be evenly spread over representation space. \cite{renart03} show this effect can be overcome by considering homeostatic mechanisms that balance excitatory drive to each neuron in the network, halting drift of bumps altogether. On the other hand, both \cite{itskov11} and \cite{hansel13} show drift can be slowed by including short term facilitation in the network. Rather than exploring ways to remove the effective drift introduced by spatial heterogeneity, we have shown spatial heterogeneity can improve network coding. As long as quantization error is outweighed by a reduction in diffusion error, heterogeneous networks make less overall error in recall tasks than spatially homogeneous networks.

Our model represents the space of possible oriented cues as evenly distributed in space with uniform probability of presentation, a protocol often used in experiments \citep{funahashi89,white94,goldmanrakic95,meyer11}.  Thus, we reason that the ideal covering of stimulus space by the network will have a uniform distribution. This translates into network spatial heterogeneity that is exactly periodic.  The periodicity allows for a compact derivation of the effective diffusion coefficient $D_{\rm eff}(n)$. Were the stimulus set to have an asymmetric probability distribution, we would expect the ideal spatial heterogeneity would not produce evenly spaced attractors.  In this case, approximating bump position is possible, but motion between attractors will depend on $\theta$ and will be difficult to interpret as a simple diffusion process.  Nevertheless, we expect that the specifics of spatially uneven heterogeneous coupling will significantly impact both the attractor quantization and stochastic drift across the network, and control the information transfer from stimulus to recall.  

\subsection{Mechanisms that produce structured heterogeneity}

We conceive of two main biophysical processes that could produce structured spatial heterogeneity in a working memory network. First, Hebbian plasticity may operate at locations in the network that are driven by common external cues. Such cues will consistently activate neurons of similar orientation preference, so clusters of similarly tuned cells will tend to strengthen recurrent excitation between each other \citep{goldmanrakic95,clopath10,ko13}. Recent experiments show training does increase the delay period firing rates of neurons with a preference for the encoded cue \citep{meyer11}, which may occur due to reenforcement of recurrent excitation. This mechanism would create attractors only at locations in the network that consistently receive feedforward input during training. Neurons that are never directly stimulated by cues would be deprived of continual reenforcement of their excitatory inputs, allowing broadly tuned inhibition to decrease their delay period firing \citep{wang01}. In the framework of our models, a network trained on $n$ cue locations would form $n$ attractors. Depending on the length of delay, this might not be the optimal number of attractors, but it would improve coding as compared to the network without quantization.

Second, reward-based plasticity mechanisms signaled by dopamine may supervise the reenforcement of synaptic excitation to form a network with the optimal number of attractors. Many studies have verified that dopamine can carry reward signals back to the network responsible for a correct action \citep{schultz98}. Selectively acting on specific subsets of neurons, dopamine can prompt plasticity in network architecture to improve future chances of rewards \citep{mcnab09,klingberg10}. Such supervisory mechanisms could seek an optimal architecture in the network to maximize reward yields for a fixed retention time and number of possible cues. As we have demonstrated, this resulting structured heterogeneity will improve coding, even if there is unstructured heterogeneity present \citep{wang06}.

\subsection{Relating diffusion of neural activity to behavior}

Our results (and those of many other past studies) assume that neural activity has a diffusive component.  However, how exactly neural variability drives behavioral variability is largely unknown \citep{britten96, churchland11,brunton13, haefner13}.  Psychophysical studies of spatial working memory tasks reveal that subjects typically respond with nonzero error.  In particular, \cite{ploner98} show that the midspread of memory guided saccades in humans scales sublinearly with delay time over $0.5-20$\! s.  This scaling is consistent with a diffusion of neural activity involved in the storage of memory, giving support to our modeling assumptions.  

In contrast to these data, recent psychophysiological work in rodents and humans performing decision making tasks lasting $0.5-2$\! s suggests that models with sensory noise, rather than internal diffusion, best capture these behavioral data \citep{brunton13}.  However, this study ultimately considers a two alternative forced choice task, and does not consider the storage of inputs over a large stimulus space.  When there are only two attractors in our network the diffusion coefficient is near zero, consistent with \cite{brunton13}.  %Nevertheless, our network would still need to integrate inputs (evidence) over time, a property that a two attractor network lacks. 
In addition, the timescale of tasks studied by \cite{brunton13} may not be long enough to substantially reveal the effects of internal diffusion. These differences between the diffusive nature of working memory and decision integrator networks suggest that more work needs to be done to link variability of neural and behavioral response.

\subsection{Implications for multiple object working memory}
We emphasize that we did not study network encoding of multiple object working memory. Added complications arise when several items must be remembered at once \citep{luck97}. For instance, the error made in recalling the value of a set of multiple continuous variables increases with the set size \citep{wilken04}. Recently, it has been shown that a spiking network model can recapitulate many of these set size effects \citep{wei12}. Interestingly, there is an optimal spread of pyramidal synapses that minimizes errors due to set size. However, reduction of the effects of dynamic noise on the accuracy of memories has yet to be studied. Our ideas could be extended to analyze how networks that encode multiple object memory could be made more robust, applying network quantization to the storage of multiple bump attractors. 

\bibliographystyle{jneurosci}
\bibliography{optimal}

\clearpage

\clearpage

\clearpage

\clearpage

\clearpage

\clearpage

\clearpage

\clearpage

\end{document}